\documentclass[10pt,aps,showpacs,amsmath,amssymb,prd,eqsecnum]{revtex4-2}

\usepackage{simpler-wick, bm, upgreek, textgreek, slashed, braket, mathtools, tensor, bbold,amsthm,amssymb,bbold,mathtools,tensor,natbib}
\usepackage[super]{nth}

\allowdisplaybreaks

\RequirePackage[colorlinks=true
  ,urlcolor=blue
  ,anchorcolor=blue
  ,citecolor=blue
  ,filecolor=blue
  ,linkcolor=blue
  ,menucolor=blue
  ,linktocpage=true
  ,pdfproducer=medialab
 ]{hyperref}

\begin{document}
\pagestyle{plain}

\title{Divergences in Nonpolynomial Quantum Field Theories}

\author{Galib\,\,Hoq}
\email{gxh368@miami.edu}
\affiliation{Department of Physics, University of Miami, Coral Gables, FL 33146}

\author{Thomas\,\,Curtright}
\email{curtright@miami.edu}
\affiliation{Department of Physics, University of Miami, Coral Gables, FL 33146}

\date{\today}

\begin{abstract}
   The structure of divergences in several nonpolynomial scalar quantum field theories are investigated. Various \(n\)-point Green functions for nonpolynomial theories are calculated and their renormalization is analyzed, with special regard for the nonpolynomial \(\left|\phi\right|^{3}\) theory in \(4\) spacetime dimensions. 
\end{abstract}

\maketitle
\thispagestyle{plain}

\section{Introduction}

Scalar field self-interactions are limited in form, in four-dimensional
spacetime (4D), by the requirements of vacuum stability and
renormalizability.  As a consequence of these two requirements, most models
only involve polynomial scalar field interactions with at most quartic
powers of the fields.  Nevertheless, around 60 years ago there were
intriguing proposals, by Efimov \cite{Efimov1963} \cite{Efimov1964} \cite{Efimov1965}, by Fradkin \cite{Fradkin1963},
and by Delbourgo, Salam, and Strathdee \cite{Salam1970} \cite{Salam1971}, that non-polynomial
interactions could perhaps lead to interesting and non-trivial
renormalizable quantum field theories involving scalar fields, even in 4D,
albeit at considerable computational expense.  Other non-polynomial models
have been considered in contemporary studies \cite{Shaposhnikov2023} \cite{Bender1987} \cite{Bender1988}, especially by
proponents of PT-symmetric theories \cite{Bender2018} \cite{Bender2021}, but again at considerable
computational expense.  

More recently \cite{preprint}, we proposed an approach to non-polynomial scalar
field self-interactions in a very tractable framework involving specific
analytic functions of \(\phi\), with computations more manageable than those
encountered in previous work.  In particular, we considered theories with \( %
V\left( \phi \right) =\kappa \left( \phi ^{2}\right) ^{\nu }\)  for generic
real  \(\nu\), where the positive principal \(\nu\)th root is understood.  We
defined such interactions as linear combinations of exponential
interactions, i.e. as Gaussians integrated over contours in the complex
plane.  We were able to obtain perturbative results for each fixed order in 
\(\kappa\) that were exact analytic functions of \(\nu\).  

In this paper we pursue this approach to investigate ultraviolet divergences
for these and related theories, where the latter involve non-Gaussian
exponentials linear in the fields (i.e. linear combinations of complex
Liouville models) that more readily incorporate interactions of the form \(%
V\left( \phi \right) =\kappa /\phi ^{\nu }\).  
\section{Outline}
The paper is organized as follows: in section 3, we show how the Schwinger-Dyson equations can be derived from the Feynman path integral. This material is well-known \cite{schwartzbook} \cite{peskinschroeder}, but we include it here in order to consolidate notation and for convenient reference. In section 4, we give the main results of the paper. More specifically, we show how power-counting can be done for nonpolynomial scalar quantuam field theories (QFTs) and give explicit formulae for the superficial degree of divergence \(D\) for these theories. In section \(5\), we calculate some \(n\)-point Green functions for various nonpolynomial theories and compare our results with those obtained by workers in PT-symmetric QFTs. In section 6 we discuss renormalization as it pertains to nonpolynomial QFTs, especially the \(\left|\phi\right|^{3}\) theory.  In the appendix, we give some of the background from \cite{preprint} and collect some technical mathematical results for easy reference. Let us now state some of our notations and conventions.  
\begin{equation}
  \label{scalarfeynmanpropagator}
  \Delta_{xy} \equiv \Delta\left(x-y\right) = \int \frac{d^{N}p}{\left(2\pi\right)^{N}}\, \frac{i\,e^{-ip\left(x-y\right)}}{p^{2}-m^{2}+i\epsilon}
\end{equation}
will often be used to denote the scalar Feynman propagator.  We will use \(\Delta(0)\) and \(\Delta\) for \(\Delta\left(y-y\right) \equiv \left\langle \phi^{2}(0) \right\rangle \).  \(\delta^{(N)}_{ij}\) will be used for the \(N\)-dimensional Dirac Delta function \(\delta^{(N)}(i-j)\). \(\phi_{x},J_{x}\) will be used to write \(\phi(x), J(x)\).  The shorthand \(\left\langle \cdots\right\rangle\) will be used for the vacuum expectation values(VEV) in the interacting theory \( \bra{\Omega} T\left\{\cdots\right\}\ket{\Omega}\) and likewise \(\left\langle \cdots\right\rangle_{0}\) for the free theory VEVs \( \bra{0} T\left\{\cdots\right\}\ket{0}\). The following analytical continuation of the double factorial will be used 
\begin{equation}
  z!! = \left(\frac{1}{\sqrt{2}}\right)^{z+1} \frac{\Gamma\left(z+2\right)}{\Gamma\left(\frac{z+3}{2}\right)} .
\end{equation}
\section{Schwinger-Dyson Equations}
In order to avoid cumbersome notation we will often use \(K\) instead of \(\left(\Box +m^2\right)\) when employing the Schwinger-Dyson equations for perturbation theory, However, the final result will be written using \(\left(\Box +m^2\right)\). For a scalar quantum field theory(QFT) we have the formal Lagrangian 
\begin{equation}
  \mathcal{L} = -\frac{1}{2}\phi\left(\,\Box+m^{2}\right)\phi + \mathcal{L}_{\text{int}}
\end{equation}
where \(\mathcal{L}_{\text{int}}\) is the interaction term (for example \(\mathcal{L}_{\text{int}}=-\frac{\kappa}{4!} \phi^{4}\) in \(\phi^{4}\) theory). The quantum theory can be defined using the Feynman path integral 
\begin{equation}
\label{vacuumgenerating1}
  Z[J] \equiv \int D\Phi \, \exp \left(i\left\{\int d^{N}y \,  \left[-\frac{1}{2}\phi\left(\Box+m^{2}\right)\phi + \mathcal{L}_{\text{int}}[\Phi] + J_{y}\phi_{y}  \right]  \right\}   \right) 
\end{equation}
The \(n\)-point Green functions of an interacting quantum field theory can be calculated using 
\begin{equation}
\left\langle \phi_{1}\cdots \phi_{n} \right\rangle \equiv \frac{\int D\Phi  \, \phi_{1}\cdots \phi_{n} \, e^{iS[\Phi]} }{  \int D\Phi  \, e^{iS[\Phi]}  } \equiv \left. \, \frac{\left(-i\right)^{n}}{Z[J]} \, \frac{\partial^{n} Z[J]}{\partial J_{1}\ldots \partial J_{n}}\right|_{J=0}
\end{equation}
where 
\begin{equation}
  S[\Phi] \equiv \int d^{N}y \,  \left[-\frac{1}{2}\phi\left(\Box+m^{2}\right)\phi + \mathcal{L}_{\text{int}}[\Phi]   \right]   
\end{equation}
To prove the Schwinger-Dyson equations for the \(n\)-point Green functions we start by writing
\begin{equation}
\label{1.5sde}
\begin{aligned}
\left\langle \phi_{1}\cdots \phi_{n} \right\rangle = \left.\frac{1}{Z[J]} \int \mathcal{D}\Phi \, \phi_{1}\cdots \phi_{n}\, e^{i\int d^{N}y\, \left\{ -\frac{1}{2}\phi K \phi + \mathcal{L}_{\text{int}}[\Phi]  +J\phi \right\}}\right|_{J=0}
\end{aligned}
\end{equation}
The path integral is invariant under the translation \( \phi \rightarrow \phi + \epsilon\)
\begin{equation}
\label{1.10sde}
\begin{aligned}
\left\langle \phi_{1}\cdots \phi_{n} \right\rangle &= \frac{1}{Z[0]} \int \mathcal{D}\left[\Phi+\epsilon\right] \, \left(\phi_{1}+\epsilon_{1}\right)\cdots \left(\phi_{n}+\epsilon_{n}\right)  e^{i\int d^{N}y\, \left\{ -\frac{1}{2} \left(\phi+\epsilon\right) K \left(\phi+\epsilon\right) + \mathcal{L}_{\text{int}}[\Phi+\epsilon]  \right\}} \\
&=  \frac{1}{Z[0]} \int \mathcal{D}\Phi \, e^{i\int d^{N}y \, S[\Phi]} \biggl\{  \phi_{1}\cdots\phi_{n} +  \epsilon_{j}\,\phi_{1}\cdots \phi_{j-1} \phi_{j+1}\cdots\phi_{n}+\cdots +\epsilon_{n} \phi_{1}\cdots\phi_{n-1} + \\ 
&+i\int d^{N}x \epsilon_{x} \mathcal{L}'_{\text{int}}[\phi_{x}] \phi_{1}\cdots \phi_{n} -i\int d^{N}x\, \epsilon_{x} K \phi_{x}\phi_{1}\cdots \phi_{n} \biggr\} +O\left(\epsilon^{2}\right)
\end{aligned}
\end{equation}
Upon comparing (\ref{1.5sde}) and (\ref{1.10sde}) we get
\begin{equation}
\begin{aligned}
0 = \int d^{N}x \, \epsilon_{x} \, \int \frac{\mathcal{D}\Phi \, e^{iS[\Phi]}}{Z[0]}  \, \Biggl\{  & -iK_{x}\phi_{x}\phi_{1}\cdots \phi_{n} +i\mathcal{L}'_{\text{int}}[\phi_{x}] \phi_{1}\cdots \phi_{n} \\ 
&+ \cdots \delta^{(N)}_{xj}\phi_{1}\cdots \phi_{j-1}\phi_{j+1}\cdots \phi_{n} + \cdots +\delta^{(N)}_{x\left(n-1\right)}\phi_{1}\cdots \phi_{n}                 \Biggr\}
\end{aligned}
\end{equation}
As this must be true for any \(\epsilon_{x}\) we can say
\begin{equation}
\begin{aligned}
0 =  \int \frac{\mathcal{D}\Phi \, e^{iS[\Phi]}}{Z[0]}  \, \Biggl\{  & -iK_{x}\phi_{x}\phi_{1}\cdots \phi_{n} +i\mathcal{L}'_{\text{int}}[\phi_{x}] \phi_{1}\cdots \phi_{n} \\ 
&+ \cdots \delta^{(N)}_{xj}\phi_{1}\cdots \phi_{j-1}\phi_{j+1}\cdots \phi_{n} + \cdots +\delta^{(N)}_{x\left(n-1\right)}\phi_{1}\cdots \phi_{n}                 \Biggr\} \\
\end{aligned}
\end{equation}
We can now reorganize in the following way 
\begin{equation}
\begin{aligned}
\frac{1}{Z[0]} \int \mathcal{D}\Phi \, e^{iS[\Phi]}\, \left(\phi_{x}\phi_{1}\cdots \phi_{n} \right) &= \frac{1}{Z[0]} \int \mathcal{D}\Phi \, e^{iS[\Phi]}\, \left(\mathcal{L}'_{\text{int}}[\phi_{x}]\phi_{1}\cdots \phi_{n} \right) \\
&-i\frac{1}{Z[0]} \int \mathcal{D}\Phi \, e^{iS[\Phi]}\, \left(\sum_{j=1}^{n} \delta^{(N)}_{xj} \phi_{1}\cdots \phi_{j-1}\phi_{j+1}\cdots \phi_{n} \right)
\end{aligned}
\end{equation}
Finally, we have
\begin{equation}
\boxed{\begin{aligned}
(\Box_{x} +m^{2})\left \langle \phi_{x}\phi_{1}\cdots \phi_{n}  \right \rangle = \left \langle \mathcal{L}'_{\text{int}}[\phi_{x}] \,\phi_{1}\cdots \phi_{n} \right \rangle - i\sum_{j=1}^{n} \delta^{(N)}_{xj} \left \langle \phi_{1}\cdots \phi_{j-1}\,\phi_{j+1} \cdots \phi_{n}  \right \rangle 
\end{aligned}}
\end{equation}
where we have again set \(K = \left(\Box_{x} +m^{2}\right)\). These are the \emph{Schwinger-Dyson equations} for the \(n\)-point Green functions. 
\section{Power Counting}

\subsection{Polynomial Scalar Field Theory}

For a polynomial scalar quantum field theory the usual Lagrangian is 
\begin{equation}
\label{lagrangianscalar}
\mathcal{L} =  -\frac{1}{2}\phi\left(\Box +m^{2}\right)\phi -  \frac{\kappa}{n!} \phi^{n}.
\end{equation}
For a scalar theory, a typical Feynman diagram (modulo numerical factors) is of the form
\begin{equation}
  \sim \int \frac{d^{N}p_{1}\cdots d^{N}p_{L}}{\left(p^{2}_{1}-m^{2}\right)\cdots \left(p^{2}_{L}-m^{2}\right)}
\end{equation}
where \(L\) is the number of loops. \(L\) is related to the number of internal Feynman propagators \(I\) and the number of vertices (or equivalently the order of perturbation) V via
\begin{equation}
\label{eulerpower}
L= I-V+1  
\end{equation}
The superficial degree of divergence \(D\) is defined as 
\begin{equation}
   D \equiv \left(\text{powers of momenta in numerator} \right)-  \left(\text{powers of momenta in denominator}\right) 
\end{equation}
Given the form of the scalar Feynman propagator (\ref{scalarfeynmanpropagator}) it is clear that in \(N\) spacetime dimensions for a scalar QFT we will have
\begin{equation}
\label{superficial2}
  D = NL-2I
\end{equation}
For polynomial scalar theories it is known \cite{peskinschroeder} that \(V\) is related to \(I\) and the number of external lines, \(E\), through
\begin{equation}
\label{powerinf}
nV= E+2I
\end{equation}
Substituting (\ref{eulerpower}) and (\ref{powerinf}) in (\ref{superficial2}) we get the following formula for the superficial degree of divergence \(D\) for any scalar polynomial quantum field theory

\begin{equation}
\label{superficial1}
 D = N + \left[n\left(\frac{N-2}{2}\right)-N\right]V-\left(\frac{N-2}{2}\right)E
\end{equation}

\subsection{Nonpolynomial Scalar Quantum Field Theories}
The main nonpolynomial scalar field theory investigated in \cite{preprint} had an interaction Lagrangian of the form 
\begin{equation}
\label{lagrangian1}
  \mathcal{L}_{\text{int}} = \frac{\kappa}{\Gamma\left(2\nu+1\right)}\left(\phi^{2}\right)^{\nu} \quad \left(\text{Re }\nu >0\right)
\end{equation}
It is not Apriori clear how a relation like  (\ref{powerinf}) could be derived.  Some of our main results in this paper will include formulae like (\ref{powerinf}) for various nonpolynomial theories. In \(\phi^{4}\) theory, for the \(2\)-point function at second order (so \(E=2\) and \(V=2\)), we get diagrams like 
\(-\kappa^{2}\frac{ \Delta^{3}_{xy}\Delta_{x1}\,\Delta_{y2}}{6}\). We note that there are total number of \(5\) propagators here (we are counting both internal propagators and external lines here as propagators). Let us denote by \(T\) the total sum of powers of the propagators at a fixed order of perturbation and for a given number of external lines. It is straighforward to see that for a given number of vertices and external lines, \(T\) will be constant for all the relevant diagrams of that order. This follows from dimensional analysis, since a particular diagram consists of the following arrangement 
\begin{equation}
 \kappa^{V} \prod_{j=1}^{V} \int d^{N}y_{j} \left\langle f(\Phi)\right\rangle_{0}
\end{equation} where \(\kappa\) is the coupling constant, \(y_{j}\) denotes internal points, and \(f(\Phi)\) is some function of the scalar quantum fields \(\phi_{1},\phi_{2},\ldots,\phi_{E}\). Since, \(\kappa\), \(d^{N}y_{j}\), and the scalar Feynman propagator \(\Delta_{ij}\) are all dimensional quantities, \(T\) must be the same in order for each diagram (at fixed values of \(V\) and \(E\)) to be dimensionally consistent.  We note that \(T\) is related to \(E\) and \(I\) via
\begin{equation}
\label{power1}
T = E+I
\end{equation}
We will now show that for theories with the Lagrangian (\ref{lagrangian1}) 
\begin{equation}
  T= V\nu + \frac{E}{2}
\end{equation}
We will divide the proof into two separate parts.
\begin{proof}
\textbf{Part (a)} \( \left(V \leq E\right)   \). Any \(n\)-point Green function can be written as
\begin{equation}
\label{begin1}
 \left \langle \phi_{1} \phi_{2}\cdots \phi_{E} \right \rangle = \int d^{N}y_{1} \delta^{(N)}_{11'} \left \langle \phi_{y_1} \phi_{x_2}\cdots \phi_{x_E} \right \rangle  
\end{equation}
Using \(   \delta^{(N)}_{11'}=i\,\Box_{y_{1}} \Delta_{11'} \) and integration by parts in (\ref{begin1}) we get 
\begin{equation}
 \left \langle  \phi_{1} \phi_{2}\cdots \phi_{E} \right \rangle  = i\int d^{N}y_{1}\, \Delta_{11'}\, \Box_{y_{1}} \left \langle \phi_{y_1} \phi_{x_2}\cdots \phi_{x_E} \right \rangle  
\end{equation}
We now use the Schwinger-Dyson equations on \(\Box_{y_{1}} \left \langle \phi_{y_1} \phi_{x_2}\cdots \phi_{x_E} \right \rangle    \) to get 
\begin{align}
\label{sd14}
  \left \langle  \phi_{1} \cdots \phi_{E} \right \rangle = i\int d^{N}y_{1} \,\kappa\frac{i\Gamma\left(1+\nu\right)}{2\pi}\oint_{C} ds_{1}\, &\left(-s_{1}\right)^{-1-\nu} \left(-s_{1}\right)  \,\Delta_{11'} \nonumber \\
  & \biggl\{  \left \langle e^{-s_{1}\phi^{2}_{y_1}} \phi_{y_{1}} \phi_{x_2}\ldots \phi_{x_E} \right \rangle              -i\sum\left(\cdots\right) \biggr\} 
\end{align}
where \(i\sum\left(\cdots\right)\) denotes the correction terms from using the Schwinger-Dyson equations. As we mentioned previously, the quantity \(T\) is the same for each diagram for fixed \(E\) and \(V\). Hence, for simplicity we can focus on a single term at a particular order(fixed values of \(E\) and \(V\)). Even numerical factors and signs are not of importance here. Hence, we can rewrite (\ref{sd14}) in the following way
\begin{equation}
  \left \langle \phi_{1} \cdots \phi_{E} \right \rangle \sim \kappa \int d^{N}y_{1} \, \oint_{C} ds_{1}\, \left(-s_{1}\right)^{-1-\nu} \left(-s_{i}\right)  \,\Delta\,    \left \langle e^{-s_{1}\phi^{2}_{y_1}} \phi_{y_{1}} \phi_{x_{2}}\cdots \phi_{x_E} \right \rangle             
\end{equation}
We are suppressing the indices of the propagators since we only care about the total sum of the powers of the propagators. It can now be seen that continuing the iteration of the Schwinger-Dyson equations will lead us to 
\begin{align}
\label{sd15}
  \left \langle \phi_{1} \cdots \phi_{E} \right \rangle \sim \kappa^{V} \prod_{i=1}^{V} \int d^{N}y_{i} \oint_{C} ds_{i}\, \left(-s_{i}\right)^{-1-\nu} \left(-s_{i}\right)  \,\Delta^{V}\, \times  \left \langle e^{-s_{1}\phi^{2}_{y_1}} \phi_{y_{1}} \,e^{-s_{2}\phi^{2}_{y_2}} \,\phi_{y_{2}}\cdots e^{-s_{V}\phi^{2}_{y_V}} \,\phi_{y_{V}} \, \phi_{x_{1}}  \cdots   \phi_{x_{E-V}} \right \rangle_{0}
\end{align}
where we have relabled the external points \(x_{i}\) to indicate the total number of terms remaining. We now use (\ref{hankel1}) and (\ref{rep1}) in (\ref{sd15}) to write 
\begin{align}
\label{sd16}
  \left \langle \phi_{1}\cdots \phi_{E} \right \rangle \sim \kappa^{V} \prod_{i=1}^{V} \int d^{N}y_{i} \, \oint_{C} &\frac{ ds_{i}\, \left(-s_{i}\right)^{-1-\nu} \left(-s_{i}\right)}{\sqrt{s_{i}}}  \, \,\int_{-\infty}^{\infty} d\sigma_{i} \, \exp\left(-\sum_{k=1}^{V} \frac{\sigma^{2}_{k}}{4s_{k}}\right) \prod_{j=1}^{E-V} \prod_{l=1}^{V}   \oint_{|\tau|=1} \,\oint_{|\lambda|=1}\, \frac{d\tau_{j}}{\tau^{2}_{j}}\,\, \frac{d\lambda_{l}}{\lambda^{2}_{l}}  \nonumber \\  
  &\Delta^{V} \,  \left \langle e^{i\sigma_{1}\phi_{y_1}}\cdots e^{i\sigma_{V}\phi_{y_V}} \, e^{i\lambda_{1}\phi_{y_{1}}}\cdots  e^{i\lambda_{V}\phi_{y_{V}}}\,  e^{i\tau_{1}\phi_{x_{1}}} \cdots  e^{i\tau_{E-V}\phi_{x_{E-V}}} \right \rangle_{0} 
\end{align}
We note at this point that all the exponential terms are \emph{dimensionless}. Hence, the entire VEV is also dimensionless. We now remark that once all the \(\sigma,\tau,\lambda\) and \(s\) integrations are carried out, no factors of \(\sigma,\tau,\lambda\) or \(s\) will remain. This means that apart from \(\kappa\) and \(d^{N}y_{i}\) the propagators will be left as the only \emph{dimensional} quantities. The relevant dimensional quantities(written in terms of mass dimensions) are 
\begin{equation}
\label{massdim1}
\begin{aligned}
&-[\Delta]= [ds] = [s] =-\left(N-2\right) \\
&  [d\sigma]=[\sigma] = [d\tau] = [\tau] = [d\lambda] = [\lambda] = -\frac{1}{2}\left(N-2\right)  
\end{aligned}
\end{equation}
Therefore, the number of propagators, \(\eta\), produced as a result of doing all the integrations, can be calculated using 
\begin{equation}
\eta\left(N-2\right) = \left(N-2\right) V\, \left(\nu-\frac{1}{2}\right)  + \frac{1}{2} \left(N-2\right) \left(E-V\right) \rightarrow \eta = V\nu +\frac{E}{2} -V  
\end{equation}
We recall that in (\ref{sd16}) we already had \(V\) propagators. So the total sum of the powers of all the propagators, \(T\) , is given by  
\begin{equation}
   T = \eta +V = \left(V\nu +\frac{E}{2} -V \right) +V= V\nu +\frac{E}{2} 
\end{equation}
which is precisely what we wanted to prove. 

\textbf{Part (b)} \( \left(V > E\right)  \). Using the Schwinger-Dyson equations we get 
\begin{equation}
\label{sd17}
\begin{aligned}
  \left \langle \phi_{1} \phi_{2}\cdots \phi_{E} \right \rangle &\sim \kappa^{E}\, \prod_{i=1}^{E} d^{N}y_{i} \oint_{C} ds_{i}\, \left(-s_{i}\right)^{-1-\nu} \left(-s_{i}\right)  \,\Delta^{E}\, \times   \,\left \langle e^{-s_{1}\phi^{2}_{y_1}} \phi_{y_{1}} \,e^{-s_{2}\phi^{2}_{y_2}} \,\phi_{y_{2}}\cdots  e^{-s_{E}\phi^{2}_{y_E}} \,\phi_{y_{E}} \right \rangle \\
&\sim \kappa^{V} \prod_{i=1}^{V} \int d^{N}y_{i}\, \oint_{C} ds_{i}\, \left(-s_{i}\right)^{-1-\nu}\,\left(-s_{i}\right) \times \\
  & \Delta^{V} \left \langle e^{-s_{1}\phi^{2}_{y_1}} \phi_{y_{1}} \,e^{-s_{2}\phi^{2}_{y_2}} \,\phi_{y_{2}}\cdots  e^{-s_{E-1}\phi^{2}_{y_E-1}} \, e^{-s_{E}\phi^{2}_{y_{E}}} \cdots  e^{-s_{V-1}\phi^{2}_{y_{V-1}}} e^{-s_{V}\phi^{2}_{y_{V}}}  \,\phi_{y_{V}} \right \rangle_{0} \\
  &\sim \kappa^{V} \prod_{i=1}^{V} \int d^{N}y_{i}\, \oint_{C} ds_{i}\, \left(-s_{i}\right)^{-1-\nu}\,\left(-s_{i}\right)  \prod_{j=1}^{E-1} \oint_{|\lambda|=1}  \, \frac{d\lambda_{j}}{\lambda^{2}_{j}} \,  \oint_{|\lambda|=1}\,\frac{d\lambda_{V}}{\lambda^{2}_{V}} \times \\
& \Delta^{V} \left \langle e^{-s_{1}\phi^{2}_{y_1}}  \,e^{-s_{2}\phi^{2}_{y_2}} \cdots  e^{-s_{E-1}\phi^{2}_{y_E-1}} \, e^{-s_{E}\phi^{2}_{y_{E}}} \cdots  e^{-s_{V}\phi^{2}_{y_{V}}}   \, e^{i\lambda_{1}\phi_{y_{1}}} \cdots e^{i\lambda_{E-1}\phi_{y_{E-1}}} \, e^{i\lambda_{V}\phi_{y_{V}}}\right \rangle_{0} 
\end{aligned}
\end{equation}
Using the method from part (\(a\)), 
\begin{equation}
\begin{aligned}
  \eta\left(N-2\right) = \left(N-2\right)V\,\left(\nu-1\right)+ \frac{1}{2}\left(N-2\right)(E) \rightarrow  \eta = V\nu -V + \frac{E}{2}
 \end{aligned}
\end{equation}
Since we already had \(V\) propagators in (\ref{sd17}), \(T\) is given by 
\begin{equation}
\begin{aligned}
  T &= \eta + V =\left(V\nu -V + \frac{E}{2}\right) + V =V\nu + \frac{E}{2} \\
\end{aligned}
\end{equation}
Therefore \(T=V\nu+\frac{E}{2}\) is true for both \(V \leq E\) and \(V > E\). We can use (\ref{power1}) to get
\begin{equation}
\label{mainresult4}
\begin{aligned}
  E+I = V\nu+\frac{E}{2} \rightarrow \boxed{\left(2\nu\right)V=E+2I} 
\end{aligned}
\end{equation}
Using (\ref{mainresult4}) and (\ref{eulerpower}) in (\ref{superficial2}) the superficial degree of divergence \(D\) for the theory with Lagrangian given by (\ref{lagrangian1}) is
\begin{equation}
\label{mainresult8}
\boxed{D = N +  \biggl[\nu\left(N-2\right) -N \biggr]V -\left(\frac{N-2}{2}\right)E  } 
\end{equation}
The degree of divergence \(D\) for this theory is a simple extension of the usual one for polynomial theories (\ref{superficial1}), with \(n  \rightarrow \left(2\nu\right)  \). 
\end{proof}
\subsection{Inverse-power Nonpolynomial Scalar Field Theory}
We now look at nonpolynomial theories of the ``inverse-power'' form 
\begin{equation}
\label{lagrangian2}
\mathcal{L}_{\text{int}} = \frac{\kappa}{\phi^{\nu}} \quad \left(\text{Re }\nu >0\right)
\end{equation}
For theories with Lagrangians (\ref{lagrangian2}), we will use (\ref{rep2}).

\begin{proof} 

\textbf{Part (a)} \( \left(V \leq E\right)\). Using  (\ref{rep2}) and the Schwinger-Dyson equations we get 
\begin{equation}
\label{sd21}
\begin{aligned}
  \left \langle \phi_{1} \cdots \phi_{E} \right \rangle &\sim \kappa^{V} \prod_{i=1}^{V} \int d^{N}y_{i}\, \int_{0}^{\infty} du_{i}\, u^{\nu-1}_{i}\,\left(-u_{i}\right) \, \Delta^{V} \left \langle e^{-u_{1}\phi_{y_{1}}}\cdots e^{-u_{V}\phi_{y_{V}}} \, \phi_{x_{1}} \cdots \phi_{x_{E-V}} \right \rangle_{0} 
\end{aligned}
\end{equation}
If we make use of (\ref{rep1}), (\ref{sd21}) can be written as 
\begin{equation}
\label{sd22}
\begin{aligned}
\left \langle \phi_{1} \cdots \phi_{E} \right \rangle &\sim \kappa^{V} \prod_{i=1}^{V} \int d^{N}y_{i}\, \int_{0}^{\infty} du_{i}\, u^{\nu-1}_{i}\,\left(-u_{i}\right) \times \\
&\prod_{j=1}^{E-V} \int_{-\infty}^{\infty} \frac{d\lambda_{j}}{\lambda^{2}_{j}} \,  \Delta^{V} \,  \left \langle e^{-u_{1}\phi_{y_{1}}}\cdots e^{-u_{V}\phi_{y_{V}}} \, e^{i\lambda\phi_{x_{1}}} \cdots e^{i\lambda\phi_{x_{E-V}}} \right \rangle_{0} 
\end{aligned}
\end{equation}
The method from the previous subsection can be used here to get
\begin{equation}
\begin{aligned}
   \eta\left(N-2\right) =  -\frac{1}{2}\left(N-2\right) V\left(\nu+1\right) + \frac{1}{2}\left(N-2\right) \left(E-V\right) \rightarrow \eta = \frac{V\left(-\nu\right)+E}{2}-V 
\end{aligned}
\end{equation}
We already had \(V\) number of propagators in (\ref{sd22}) 
\begin{equation}
\begin{aligned}
  T = \eta+V =  \left\{\frac{V\left(-\nu\right)+E}{2}-V\right\} + V = \frac{V\left(-\nu\right)+E}{2} .
\end{aligned}
\end{equation}
\textbf{Part (b)} \( \left(V >E\right)\). Using the Schwinger-Dyson equations one finds

\begin{equation}
\label{sd23}
\begin{aligned}
  \left \langle \phi_{1} \cdots \phi_{E} \right \rangle &\sim \kappa^{E} \, \prod_{i=1}^{E} \, \int d^{N}y_{i} \, \Delta^{E} \left \langle \frac{1}{\phi^{\nu+1}_{y_{1}}} \cdots \frac{1}{\phi^{\nu+1}_{y_{E}}}   \right \rangle \\
&\sim \kappa^{E}\, \prod_{i=1}^{E} \, \int d^{N}y_{i} \, \Delta^{E} \left \langle \frac{1}{\phi^{\nu+1}_{y_{1}}} \,\frac{1}{\phi^{\nu+1}_{y_{2}}} \cdots \frac{1}{\phi^{\nu+1}_{y_{E-1}}}  \frac{1}{\phi^{\nu+2}_{y_{E}}}  \, \phi_{y_{E}}  \right \rangle 
\end{aligned}
\end{equation}
where we have made use of \(\frac{\partial \mathcal{L}_{\text{int}}}{\partial \phi} =- \kappa  \,\nu \, \frac{1}{\phi^{\nu+1}} \) and \(  \frac{1}{\phi^{\nu+1}_{E}} = \frac{1}{\phi^{\nu+2}_{E}} \, \phi_{y_{E}}\).

Iterating the Schwinger-Dyson equations now gives us 
\begin{equation}
\label{sd24}
\begin{aligned}
\left \langle \phi_{1} \cdots \phi_{E} \right \rangle &\sim \kappa^{V}\, \prod_{i=1}^{V} \, \int d^{N}y_{i} \, \Delta^{V}\,  \left \langle \frac{1}{\phi^{\nu+1}_{y_{1}}} \,\frac{1}{\phi^{\nu+1}_{y_{2}}} \cdots \frac{1}{\phi^{\nu+1}_{y_{E-1}}}\,  \frac{1}{\phi^{\nu+2}_{y_{E}}}  \cdots \frac{1}{\phi^{\nu+2}_{y_{V-1}}} \frac{1}{\phi^{\nu+1}_{y_{V}}}   \right \rangle_{0} \\
&\sim \kappa^{V}\, \prod_{i=1}^{V} \, \int d^{N}y_{i} \, \int_{0}^{\infty} du_{i} \, \prod_{j=1}^{E-1} u^{\nu}_{j} \, \prod_{k=E}^{V-1} u^{\nu+1}_{k} \, u^{\nu}_{V}  \, \Delta^{V}\,  \left \langle e^{-u_{1}\phi_{y_1}} \cdots e^{-u_{V}\phi_{y_V}} \right \rangle_{0}  
\end{aligned}
\end{equation}
As usual we can calculate
\begin{equation}
\begin{aligned}
  \eta\left(N-2\right) = -\frac{1}{2}\left(N-2\right)\left( 2V + V\nu-E  \right) \rightarrow \eta = -V -\frac{V\nu}{2} +\frac{E}{2} 
\end{aligned}
\end{equation}
We already had \(V\) propagators in (\ref{sd24})
\begin{equation}
\label{mainresult7}
\begin{aligned}
    T = \eta +V =  \left(-V -\frac{V\nu}{2} +\frac{E}{2}\right) +V= \frac{V\left(-\nu\right)+E}{2} 
\end{aligned}
\end{equation}
which we now see is true for both \(V > E\) and \(V \leq E\). Using (\ref{mainresult7}) and (\ref{power1}) we  have 
\begin{equation}
\label{mainresult6}
\boxed{V\left(-\nu\right)= E+2I}  
\end{equation}
Using (\ref{mainresult6}) and (\ref{eulerpower}) in (\ref{superficial2}), the superficial degree of divergence \(D\) for the ``inverse-power'' theory is given by
\begin{equation}
\boxed{D = N -  \biggl[\nu\left(\frac{N-2}{2}\right) +N \biggr]V -\left(\frac{N-2}{2}\right)E  } 
\end{equation}
We note that the only difference here from the analogous formula for polynomial scalar theories (\ref{superficial1}) is that \(n \rightarrow -\nu\).

\end{proof}

\subsection{Hyperbolic Cosine Nonpolynomial Theory}

\begin{equation}
\label{lagrangian3}
\mathcal{L}_{\text{int}} = M^{4} \, \left[\cosh\left[\sqrt{g_{0}}\, \frac{\phi}{M}  \right]-1\right]
\end{equation}
where  \(g_{0}\) is the coupling constant and \(M\) is the mass. We now wish to use our methods to prove the following formula given in \cite{Santonocito2023} for the superficial degree of divergence 
\begin{equation}
  D = 2P- E-4V +4
\end{equation}
where \(P\) is the power of \(g_{0}\) at a given order of perturbation. 

\begin{proof}
\textbf{Part(a)}  \( \left(V \leq E\right) \) Using the Schwinger-Dyson equations as before
\begin{equation}
\begin{aligned}
\left \langle \phi_{1} \cdots \phi_{E} \right \rangle & \sim \kappa^{V}\,\prod_{i=1}^{V} \int d^{N}y_{i}\, \prod_{j=1}^{E-V} \oint_{|\lambda_{i}|=1} \frac{d\lambda_{i}}{\lambda^{2}_{i}}\, \Delta^{V} \times \\
&   \left\langle  \sinh\left[\sqrt{g_{0}}\, \frac{\phi_{y_1}}{M}  \right]  \cdots \sinh\left[\sqrt{g_{0}}\, \frac{\phi_{y_V}}{M}  \right] \, e^{i\lambda_{1}\phi_{x_1}}\cdots e^{i\lambda_{E-V}\phi_{x_{E-V}}} \right\rangle_{0}
\end{aligned}
\end{equation}
As hyperbolic sines are also dimensionless, we now find

\begin{equation}
\begin{aligned}
  \eta\left(N-2\right) = \left(-1\right)^{2} \, \left(\frac{E-V}{2}\right) \left(\frac{N-2}{2}\right) \rightarrow \eta = \frac{E}{2} - \frac{V}{2}
\end{aligned}
\end{equation}
and 
\begin{equation}
\begin{aligned}
  T = \eta + V  = \frac{E}{2} - \frac{V}{2} + V = \frac{E+V}{2}.
\end{aligned}
\end{equation}
\textbf{Part(b)} \(\left(V >E\right)\). The Schwinger-Dyson equations give us 
\begin{equation}
\label{sd26}
\begin{aligned}
  \left \langle \phi_{1} \cdots \phi_{E} \right \rangle &\sim \kappa^{E}\,\prod_{i=1}^{E} \int d^{N}y_{i}\, \Delta^{E}\,\left\langle  \sinh\left[\sqrt{g_{0}}\, \frac{\phi_{y_1}}{M}  \right]  \cdots \sinh\left[\sqrt{g_{0}}\, \frac{\phi_{y_E}}{M}  \right] \right\rangle \\
  &\sim  \kappa^{E}\,\prod_{i=1}^{E} \int d^{N}y_{i}\, \Delta^{E}\,\left\langle  \sinh\left[\sqrt{g_{0}}\, \frac{\phi_{y_1}}{M}  \right]  \cdots \sinh\left[\sqrt{g_{0}}\, \frac{\phi_{y_E}}{M}  \right] \frac{\phi_{y_{E}}}{\phi_{y_{E}}} \right\rangle \\
  &\sim \kappa^{V}\,\prod_{i=1}^{V} \int d^{N}y_{i}\,  \prod_{j=E}^{V-1} \int_{0}^{\infty} du_{j}  \times \\
  & \Delta^{V}\,\left\langle  \sinh\left[\sqrt{g_{0}}\, \frac{\phi_{y_1}}{M}  \right]  \cdots \sinh\left[\sqrt{g_{0}}\, \frac{\phi_{y_E}}{M}  \right] \,e^{-u_{E}\phi_{y_{E}}}\,e^{-u_{E+1}\phi_{y_{E+1}}}  \cdots e^{-u\phi_{y_{V-1}}}\right\rangle_{0}
\end{aligned}
\end{equation}
We now have 
\begin{equation}
\label{finis}
\begin{aligned}
\eta\left(N-2\right) = -\frac{1}{2}\left(N-2\right)\left(V-E\right) \rightarrow \eta = \frac{E}{2}-\frac{V}{2} \quad \text{and } \quad T =\eta+V = \frac{V+E}{2}
\end{aligned}
\end{equation}
Hence, this holds for both \(V \leq E\) and \(V > E\). Using (\ref{power1}) and (\ref{finis}) 
\begin{equation}
\label{finis2}
  I = \frac{V-E}{2}
\end{equation}
Given the form of the interaction Lagrangian in (\ref{lagrangian3}) it is clear that \(P\) is related to \(V\) through 
\begin{equation}
\label{power3}
P = \frac{V}{2} \rightarrow 2P= V 
\end{equation}
Using (\ref{power3}), (\ref{eulerpower}), and (\ref{finis2}) in (\ref{superficial2}), we can say that the superficial degree of divergence \(D\) will be
\begin{equation}
\label{hyperbolicfinalresult1}
\boxed{D =2P-E -4V+4}  
\end{equation}

\end{proof}

\section{Green Functions}

\subsection{2-point Green Function}
Our Lagrangian is 
\begin{equation}
\label{lagrangian4}
\mathcal{L} =   \frac{1}{2}\left(\partial^{\mu} \phi\right)\left(\partial_{\mu}\phi\right) -\frac{1}{2}m^{2}\phi^{2} -\frac{\kappa}{\Gamma\left(2\nu+1\right)}\left(\phi^{2}\right)^{\nu}
\end{equation}
(\ref{hankel1}) allows us to take the derivative as follows
\begin{equation}
\label{1.1}
\mathcal{L}'_{\text{int}} = - \frac{\kappa}{\Gamma\left(2\nu +1\right)} \, \frac{i\Gamma\left(1+\nu\right)}{2\pi} \oint_{C} ds\, \left(-s\right)^{-1-\nu}\left(-2s\right)\,\phi_{y} \, e^{-s\phi^{2}_{y}}
\end{equation}
By using the Schwinger-Dyson equations
\begin{equation}
\label{1.2}
\begin{aligned}
  \left\langle \phi_{1}\phi_{2}  \right\rangle &= i \int d^{N}x\, \Delta_{x1}\,\Box_{x}  \left\langle \phi_{x}\phi_{2}  \right\rangle \\
  &= \Delta_{12} - \frac{i\kappa}{\Gamma\left(2\nu +1\right)}  \,  \int d^{N}y\,\Delta_{y1}\, \Delta_{y2}  \left\{  \, \frac{i\Gamma\left(1+\nu\right)}{2\pi} \oint_{C} ds\, \left(-s\right)^{-1-\nu} \, \left(-2s\right)\, \Box_{y}\left\langle e^{-s\phi^{2}_{x}}\phi_{x}\phi_{y}\right\rangle_{0}  \right\} \\
  &+O\left(\kappa^{2}\right)
\end{aligned}
\end{equation}
which will let us calculate the \(2\)-point Green function to first order. By using (\ref{appendix1})
\begin{equation}
\label{2pointfirstorder}
\begin{aligned}
  \left\langle \phi_{1}\phi_{2}  \right\rangle &= \Delta_{12} - \frac{i\kappa}{\Gamma\left(2\nu +1\right)}  \int d^{N}y\,\Delta_{y1}\, \Delta_{y2}  \left\{  \frac{i}{2\pi} \oint_{C} ds\, \frac{\left(-s\right)^{-1-\nu}\left(-2s\right)}{\left(1+2s\Delta(0)\right)^{\frac{3}{2}}} \right\} +O\left(\kappa^{2}\right)\\
  &= \Delta_{12} - \frac{i\kappa}{\Gamma\left(2\nu+1\right)} \left(\frac{2^{\nu+1}\,\nu\, \Gamma\left(\nu+\frac{1}{2}\right)}{\sqrt{\pi}}\right)\, \int d^{N}y \,\Delta^{\nu-1}(0) \, \Delta_{y1}\,\Delta_{y2} +O\left(\kappa^{2}\right)
\end{aligned}
\end{equation}
where we made use of (\ref{appendix2b}) in the last line. If we wish to calculate the second-order corrections, we can proceed mostly as before, the only additional results that we will need are
\begin{equation}
\label{1.5}
\begin{aligned}
\left \langle \phi^{2n}_{y}\right \rangle &=  i \int d^{N}x \, \Delta_{xy} \, \Box_{x} \left \langle \phi_{x}\phi^{2n-1}_{2y} \right \rangle \\
&= i \int d^{N}x \, \Delta_{xy} \, \left[ \left \langle \mathcal{L}'[\phi_{x}] \phi^{2n-1}_{y} \right \rangle_{0} -i\delta^{(N)}_{xy}\left(2n-1\right)\left \langle \phi^{2n-1}_{y}\right \rangle\right] \\
&= i \int d^{N}x \, \Delta_{xy} \, \left \langle \mathcal{L}'[\phi_{x}] \phi^{2n-1}_{y} \right \rangle_{0} + \left(2n-1\right) i\int d^{N}x \,\Delta_{xy} \,\Delta(0)\,\left \langle \mathcal{L}'[\phi_{x}] \phi^{2n-3}_{y} \right \rangle_{0} \\
&+ \left(2n-1\right)\left(2n-3\right) i\int d^{N}x \, \Delta_{xy}\Delta^{2}(0)\, \left\langle  \mathcal{L}'[\phi_{x}]  \phi^{2n-5}_{y} \right\rangle_{0} +\cdots + \Delta^{n}(0)\left(2n-1\right)\left(2n-3\right)\left(2n-5\right)\ldots 1
\end{aligned}
\end{equation}
So we have 
\begin{equation}
\label{1.6}
\begin{aligned}
&\sum_{n=0}^{\infty} \frac{\left(-s\right)^{n}}{n!} \left(2n+1\right) \left\langle \phi^{2n}_{y}\right\rangle \\
&= \sum_{n=0}^{\infty} \frac{\left(-s\right)^{n}}{n!} \sum_{m=0}^{n-1} i \int d^{N}x\,\Delta_{xy}\Delta^{m}(0) \frac{\left(2n+1\right)!!}{\left(2n-2m-1\right)!!} \left\langle\mathcal{L}'[\phi_{x}] \phi^{2n-2m-1}_{y} \right\rangle_{0} + \sum_{n=0}^{\infty} \frac{\left(-s\right)^{n}}{n!}   \left(2n+1\right)!!\, \Delta^{n}(0)
\end{aligned}
\end{equation}
where \(\mathcal{L}'[\phi_{x}]\) is given by (\ref{1.1}). By using (\ref{1.5}), (\ref{1.6}), (\ref{appendix2}), (\ref{appendix2a}), and (\ref{appendix2b}) we can show, after some lengthy algebra
\begin{equation}
\label{2pointgeneral1}
\boxed{\begin{aligned}
&\left\langle \phi_{1}\phi_{2}  \right\rangle = \Delta_{12} - \frac{i\,\kappa}{\Gamma\left(2\nu+1\right)} \, \frac{2^{\nu+1}\, \nu \, \Gamma\left(\nu+\frac{1}{2}\right)}{\sqrt{\pi}} \, \int d^{N}y\, \Delta^{\nu-1}(0)\, \Delta_{y1}\,\Delta_{y2} \\
&-\kappa^{2} \left(\frac{4^{1-\nu}}{\left[\Gamma(\nu)\right]^{2}}\right)\, \int d^{N}x\, d^{N}y\, \Delta_{xy}\,\Delta^{2\nu-2}(0)\, \Delta_{x1}\,\Delta_{y2}\,\, _2F_1\left(1-\nu ,1-\nu ;\frac{3}{2};\frac{\Delta^{2}_{xy}}{\Delta^{2}}\right) \\
& -\kappa^{2} \left[\frac{\Gamma\left(1+\nu\right)}{\Gamma\left(1+2\nu\right)} \right]^{2}\, \left[\frac{2^{1+2\nu} \left[\Gamma\left(\nu+\frac{1}{2}\right)\right]^{2}   }{\pi\, \Gamma(\nu) \, \Gamma\left(1+\nu\right)}\right] \int d^{N}x\, d^{N}y \, \Delta^{2\nu-1}(0) \, \Delta_{y1}\,\Delta_{y2}\,\,_2F_1\left(1-\nu ,-\nu ;\frac{1}{2};\frac{\Delta^{2}_{xy}}{\Delta^{2}}\right) \\
&+\kappa^{2} \left[\frac{2^{1-2\nu}}{\Gamma(\nu) \, \Gamma\left(1+\nu\right)}\right]\, \int d^{N}x\, d^{N}y \, \Delta^{2\nu-1}(0) \, \, \Delta_{y1}\,\Delta_{y2} + O\left(\kappa^{3}\right)
\end{aligned}}
\end{equation}
Setting \(\nu=2\) in (\ref{2pointgeneral1}) shows that our expression is consistent with the known result for \(\phi^{4}\) theory \cite{Ramond}. We can now also compare our results with those obtained in \cite{Bender2018} for the interaction Lagrangian 
\begin{equation}
  \frac{1}{2}\phi^{2}\left(i\phi\right)^{\epsilon} \quad \left(\epsilon \geq 0\right)
\end{equation}
If we set \(\nu \rightarrow 2+\epsilon\) and \(\frac{\kappa}{\Gamma\left(2\nu+1\right)} \rightarrow \frac{1}{2}\) our result for the \(2\)-point Green function in (\ref{2pointgeneral1}) becomes (to first order in \(\kappa\) and in Euclidean signature) 
\begin{equation}
\label{compare1}
\begin{aligned}
   \left\langle \phi_{1}\phi_{2}  \right\rangle &= \Delta_{12} -\frac{2^{\frac{\epsilon }{2}+1} \left(\frac{\epsilon }{2}+1\right) \Delta ^{\epsilon /2} \Gamma \left(\frac{\epsilon }{2}+\frac{3}{2}\right)}{\sqrt{\pi }} \int d^{N}y\,   \Delta_{y1}\,\Delta_{y2} \\
   &= \Delta_{12} - \left[  1+\frac{1}{2} \epsilon  \left(\log (\Delta )+1+\log (2)+\psi ^{(0)}\left(\frac{3}{2}\right)\right) \right]\, \int d^{N}y\,  \Delta_{y1}\,\Delta_{y2} +O\left(\epsilon ^2\right) \\
   &= \Delta_{12}-\left\{1+\epsilon \left(\frac{3}{2}- \frac{\gamma }{2}+   \frac{1}{2} \log \left[\frac{\Delta(0) }{2}\right]\right)\right\}\, \int d^{N}y\,  \Delta_{y1}\,\Delta_{y2} +O\left(\epsilon ^2\right)
\end{aligned}
 \end{equation}
If we absorb the term with no epsilon in a redefinition of the mass we get 
\begin{equation}
\left\langle \phi_{1}\phi_{2}  \right\rangle = \Delta_{12} - \epsilon \left(\frac{3}{2}- \frac{\gamma }{2}+   \frac{1}{2} \log \left[\frac{\Delta(0) }{2}\right]\right) \int d^{N}y\, \Delta_{y1}\,\Delta_{y2} +O\left(\epsilon^{2}\right)
\end{equation}
This is exactly the result obtained in \cite{Bender2018} for the two-point Green function to first order. However, our result can be expanded to \emph{all} orders in \(\epsilon\) as is clear from the \nth{1} line in (\ref{compare1}). 

\subsection{4-point Green Function}
The Schwinger-Dyson equations can be used in a way that is analogous to what we did for the \(2\)-point Green function 
\begin{equation}
  \left\langle \phi_{1}\phi_{2}\phi_{3}\phi_{4} \right\rangle = i \int d^{N}y\, \Delta_{y1} \, \Box_{y} \left\langle \phi_{y}\phi_{2}\phi_{3}\phi_{4} \right\rangle
\end{equation}
Using the results from the calculations in the \(2\)-point Green functions and (\ref{appendix2}), (\ref{appendix2a}), and (\ref{appendix2b}) we can obtain 
\begin{equation}
\label{4pointgeneral1}
\boxed{\begin{aligned}
\left \langle \phi_{1}\phi_{2}\phi_{3}\phi_{4} \right \rangle  &= \Delta_{12}\Delta_{34}+\Delta_{13}\Delta_{24} + \Delta_{14}\Delta_{23} \\
& - \frac{i\kappa}{\Gamma\left(2\nu+1\right) }\left( \frac{2^{\nu +2} \,\Gamma \left(\nu +\frac{1}{2}\right) \,\Gamma (\nu +1)}{\sqrt{\pi }\, \Gamma (\nu -1)}  \right) \int d^{N}y\, \Delta^{\nu-2}(0)\, \Delta_{y1}\,\Delta_{y2}\,\Delta_{y3}\,\Delta_{y4} \\
&- \frac{i\kappa}{\Gamma\left(2\nu+1\right) }\left(  \frac{2^{\nu +1} \,\nu  \,\Gamma \left(\nu +\frac{1}{2}\right)}{\sqrt{\pi } }  \right) \int d^{N}y\, \left\{ \Delta^{\nu-1}(0)\,\Delta_{y1}\,\Delta_{y3}\,\Delta_{24} + \text{permutations } \right\}  +O\left(\kappa^{2}\right)
\end{aligned}}
\end{equation}
We can again compare our results with those obtained for the \(4\)-point Green function in \cite{Bender2018}. Using the same method as in the previous section we get for the connected part of our \(4\)-point Green function 
\begin{equation}
\label{compare2}
\begin{aligned}
\left \langle \phi_{1}\phi_{2}\phi_{3}\phi_{4} \right \rangle_{c}  &=  -\frac{2^{\frac{\epsilon }{2}+2} \Delta ^{\frac{\epsilon }{2}-1} \Gamma \left(\frac{\epsilon }{2}+\frac{3}{2}\right) \Gamma \left(\frac{\epsilon }{2}+2\right)}{\sqrt{\pi } \Gamma \left(2 \left(\frac{\epsilon }{2}+1\right)+1\right) \Gamma \left(\frac{\epsilon }{2}\right)} \int d^{N}y\, \Delta_{y1}\,\Delta_{y2}\,\Delta_{y3}\,\Delta_{y4} \\
&= -\frac{\epsilon }{ \Delta(0) } \int d^{N}y\, \Delta_{y1}\,\Delta_{y2}\,\Delta_{y3}\,\Delta_{y4}   +O\left(\epsilon ^2\right)
\end{aligned}
\end{equation}
This is precisely what was obtained in \cite{Bender2018} for the connected four-point Green function to \(O\left(\epsilon^{2}\right)\). Just like the \(2\)-point Green function, our result for the connected \(4\)-point function can also be expanded to \emph{all} orders in \(\epsilon\) as is apparent from the \nth{1} line in (\ref{compare2}).

\section{Normal Ordering}
\subsection{Normal Ordered Lagrangians}
We know from \cite{Salam1971} \cite{Din1971} \cite{HouardTeffo1973} \cite{Ellis2016} that a general prescription for normal ordering (at least \emph{formally}) is the following
\begin{equation}
\label{normal1}
N\left\{f\left(\phi\right)\right\} \equiv e^{-\frac{\Delta(0)}{2} \frac{\delta^{2}}{\delta \phi^{2}}  } \, f\left(\phi\right)
\end{equation}
where \(f\left(\phi\right)\) is a function of scalar fields. The right-hand side of (\ref{normal1}) can be understood as an \emph{Inverse Weierstrass Transform} (see chapter 8 of \cite{WidderHirschman})
\begin{equation}
  e^{-a\frac{\delta^{2}}{\delta\phi^{2}} } \, f\left(\phi\right) \equiv \int_{-\infty}^{\infty} \frac{dk}{\sqrt{4\pi a}}\, f\left(ik\right) \, e^{-\frac{\left(k+i\phi\right)}{4a}}
\end{equation}
To see how this works. We apply it to something like \(\phi^{4}\)
\begin{equation}
\begin{aligned}
N\left\{\phi^{4}\right\}  =  e^{-\frac{\Delta(0)}{2}\frac{\delta^{2}}{\delta \phi^{2}}  } \,\phi^{4} = \phi^{4}-6\Delta(0)\phi^{2} +3\Delta^{2}(0)
\end{aligned}
\end{equation}
The vacuum expectation value (VEV) of this expression is 
\begin{equation}
\left\langle  N\left\{\phi^{4}\right\} \right\rangle_{0} = \left\langle \phi^{4}\right\rangle_{0} -6\Delta(0)\left\langle \phi^{2} \right\rangle_{0} + 3\Delta^{2}(0) \left\langle 1\right\rangle_{0} = 3\Delta^{2}(0) -6\Delta(0)\Delta(0)+3\Delta^{2}(0)\left(1\right) =0
\end{equation}
as expected. Let us now apply this to an exponential interaction Lagrangian
\begin{equation}
\begin{aligned}
N\left\{e^{i\sigma \phi}\right\} = e^{-\frac{\Delta(0)}{2}\frac{\delta^{2}}{\delta \phi^{2}}  } \, e^{i\sigma \phi} = e^{\frac{\Delta(0)}{2}\sigma^{2}} \, e^{i\sigma \phi}
\end{aligned}
\end{equation}
We can now normal order the interaction Lagrangian given in (\ref{lagrangian4}) 
\begin{equation}
\begin{aligned}
N\left\{\kappa\left(\phi^{2}\right)^{\nu}\right\} &= \kappa\frac{i\Gamma\left(1+\nu\right)}{2\pi}\oint_{C} ds\, \left(-s\right)^{-1-\nu}\, N\left\{e^{-s\phi^{2}}\right\} \\
&= \kappa\frac{i\Gamma\left(1+\nu\right)}{2\pi}\oint_{C} ds\, \left(-s\right)^{-1-\nu}\, \int_{-\infty}^{\infty} \frac{d\sigma}{\sqrt{4\pi s}}\, e^{-\frac{\sigma^{2}}{4s}} N\left\{e^{i\sigma \phi}\right\} \\
&= \kappa\frac{i\Gamma\left(1+\nu\right)}{2\pi}\oint_{C} ds\, \left(-s\right)^{-1-\nu}\, \int_{-\infty}^{\infty} \frac{d\sigma}{\sqrt{4\pi s}}\, e^{-\frac{\sigma^{2}}{4s}} \, e^{\frac{\Delta(0)\sigma^{2}}{2}}\,e^{i\sigma \phi}
\end{aligned}
\end{equation}
where we have made use of the Hankel contour expansion and (\ref{hubbard1}). 
\subsection{Green Functions}
We can now revisit our calculations for the \(2\)-point Green function, where we will now make use of the \emph{normal ordered} interaction Lagrangian. It can be seen that the only nonzero terms to \(O\left(\kappa^{2}\right)\) will be
\begin{equation}
\begin{aligned}
\left\langle\phi_{1}\phi_{2} \right\rangle = \Delta_{12} -&\left(\frac{\kappa}{\Gamma\left(2\nu+1\right)}  \right)^{2} \times \\
 &\int d^{4}y \,d^{4}x \, \Delta_{x1}\Delta_{y2} \left(\frac{i\Gamma\left(1+\nu\right)}{2\pi}\right)^{2} \left\{ \oint_{C}\oint_{C} ds\,dt\, \left(-s\right)^{-1-\nu} \,\left(-t\right)^{-1-\nu} \, 4st \left\langle N\left(e^{-s\phi^{2}_{x}}\right)N\left(e^{-t\phi^{2}_{y}}\right) \phi_{x}\phi_{y}\right\rangle_{0} \right\} 
\end{aligned}
\end{equation}
Using the methods outlined in appendix \ref{appA} and appendix \ref{appB} we can get
\begin{equation}
\label{result1}
\left\langle N\left(e^{-s\phi^{2}_{x}}\right)N\left(e^{-t\phi^{2}_{y}}\right) \phi_{x}\phi_{y}\right\rangle_{0} = \frac{\Delta_{xy}}{\left(1-4st\Delta^{2}_{xy}\right)^{\frac{3}{2}}}
\end{equation}
Therefore we have
\begin{equation}
\label{line44}
\begin{aligned}
  \left\langle\phi_{1}\phi_{2} \right\rangle = \Delta_{12} -\left(\frac{\kappa}{\Gamma\left(2\nu+1\right)}  \right)^{2}  \int d^{4}y \,d^{4}x \, \Delta_{x1}\Delta_{y2} \left(\frac{i\Gamma\left(1+\nu\right)}{2\pi}\right)^{2} \left\{ \oint_{C}\oint_{C} ds\,dt\, \left(-s\right)^{-1-\nu} \,\left(-t\right)^{-1-\nu} \, \frac{4st \Delta_{xy}}{\left(1-4st\Delta^{2}_{xy}\right)^{\frac{3}{2}}}\right\} 
\end{aligned}
\end{equation}
In order to calculate 
\begin{equation}
 \left(\frac{i\Gamma\left(1+\nu\right)}{2\pi}\right)^{2} \left\{ \oint_{C}\oint_{C} ds\,dt\, \left(-s\right)^{-1-\nu} \,\left(-t\right)^{-1-\nu} \, \frac{4st }{\left(1-4st\Delta^{2}_{xy}\right)^{\frac{3}{2}}}\right\} 
\end{equation}
We will change variables in the integration over the variable \(t \rightarrow -t\) and make use of (\ref{rep2})
\begin{equation}
\begin{aligned}
\int_{0}^{\infty} \frac{du}{\Gamma\left(\frac{3}{2}\right)} \, u^{\frac{1}{2}}\,e^{-u}\, \left(\frac{i\Gamma\left(1+\nu\right)}{2\pi}\right)^{2} \left\{ \oint_{C}\oint_{C'} ds\,dt\, \left(-s\right)^{-1-\nu} \,t^{-1-\nu} \, 4st \, e^{-s\left(4ut\Delta^{2}_{xy}\right)}\right\}
\end{aligned}
\end{equation}
where \(C'\) is the transformation of the Hankel contour after the above-mentioned change of variables. From (\ref{hankel1}) we can see that the contour integral over \(s\) can be evaluated as 
\begin{equation}
\label{line12}
\begin{aligned}
 &-4 \int_{0}^{\infty} \frac{du}{\Gamma\left(\frac{3}{2}\right)} \, u^{\frac{1}{2}}\,e^{-u}\, \left[\frac{i\Gamma\left[1+\nu\right]}{2\pi} \oint_{C'} dt\, t^{-\nu} \,\nu  \left(4ut\Delta^{2}_{xy}\right)^{\nu-1} \right] \\
 &= -\frac{4^{\nu} \Delta^{2\nu-1}_{xy}\nu}{\Gamma\left(\frac{3}{2}\right)} \int_{0}^{\infty} du\, u^{\nu-\frac{1}{2}}\,e^{-u} \left(\frac{i\Gamma\left(1+\nu\right)}{2\pi} \oint_{C'} \frac{dt}{t} \right)
\end{aligned}
\end{equation}
We know from the theory of complex variables that the contour integral in parenthesis in the last line of (\ref{line12}) will be \(2\pi i\). Thus we now have  
\begin{equation}
\begin{aligned}
  -\frac{4^{\nu} \Delta^{2\nu-1}_{xy}\nu}{\Gamma\left(\frac{3}{2}\right)} \int_{0}^{\infty} du\, u^{\nu-\frac{1}{2}}\,e^{-u} \left(2\pi i\right) &= -\frac{4^{\nu} \Delta^{2\nu-1}_{xy}\nu}{\Gamma\left(\frac{3}{2}\right)} \int_{0}^{\infty} du\, u^{\nu-\frac{1}{2}}\,e^{-u}  \, \Gamma\left(\nu+\frac{1}{2}\right) \\
  &=\frac{2^{2\nu+1}\,\nu\Gamma\left(\nu+1\right)\Gamma\left(\nu+\frac{1}{2}\right)}{\sqrt{\pi}}\,\Delta^{2\nu-1}_{xy}
\end{aligned}
\end{equation}
Putting this result back into (\ref{line44}) and simplifying, we can write the \(2\)-point Green function for the normal ordered interaction Lagrangian \(N\left\{\kappa \left(\phi^{2}\right)^{\nu}\right\}\) as 
\begin{equation}
\label{normalordergreenfunction2final}
\boxed{\begin{aligned}
\left\langle \phi_{1}\phi_{2}    \right\rangle = \Delta_{12} -\frac{\kappa^{2}}{\Gamma\left(2\nu\right)} \int d^{4}y\,d^{4}x\, \Delta^{2\nu-1}_{xy}\, \Delta_{x1}\,\Delta_{y2} +O\left(\kappa^{3}\right)
\end{aligned}}
\end{equation}
Diagramatically, the second order term in (\ref{normalordergreenfunction2final}) is  
\begin{figure}[h!]
\centering
\includegraphics[scale=0.1]{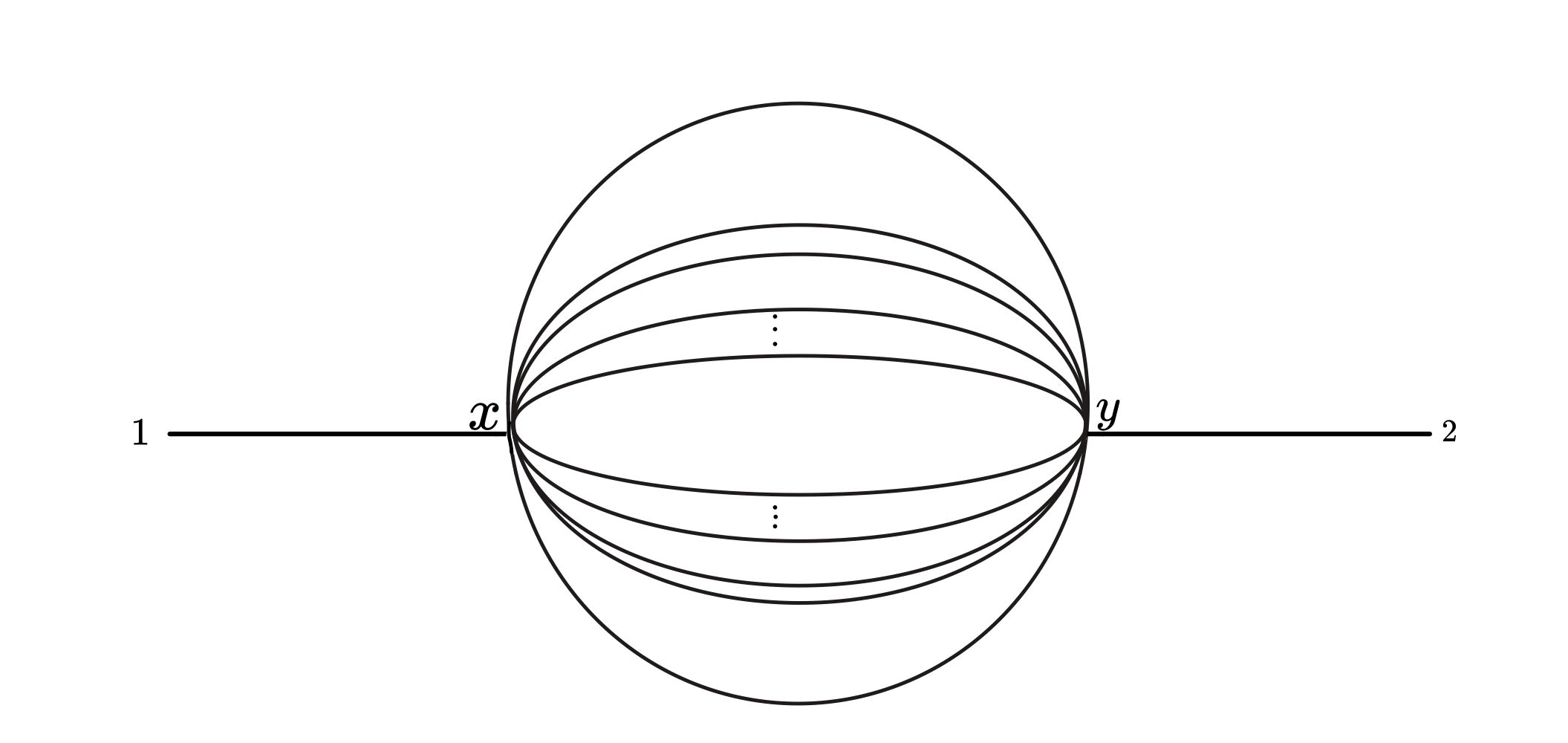}
\caption{Feynman Diagram for the \(O\left(\kappa^{2}\right)\) term.}
\end{figure}

\noindent For \(\nu=2\) we get 
\begin{equation}
\left\langle \phi_{1}\phi_{2}    \right\rangle = \Delta_{12} -\frac{\kappa^{2}}{6}\int d^{4}y\,d^{4}x\, \Delta^{3}_{xy}\,\Delta_{x1}\,\Delta_{y2}  +O\left(\kappa^{3}\right)
\end{equation}
which is exactly the so-called ``sunset diagram'' for \(\phi^{4}\) theory. For \(\nu=3\) and \(\nu=4\) we get 
\begin{equation}
\begin{aligned}
&\left\langle \phi_{1}\phi_{2}    \right\rangle = \Delta_{12} -\frac{\kappa^{2}}{120}\int d^{4}y\,d^{4}x\, \Delta^{5}_{xy}\,\Delta_{x1}\,\Delta_{y2}  +O\left(\kappa^{3}\right) \quad \left(\nu =3\right) \\
&\left\langle \phi_{1}\phi_{2}    \right\rangle = \Delta_{12} -\frac{\kappa^{2}}{5040}\int d^{4}y\,d^{4}x\, \Delta^{7}_{xy}\,\Delta_{x1}\,\Delta_{y2}  +O\left(\kappa^{3}\right) \quad \left(\nu =4\right)
\end{aligned}
\end{equation}
\section{Renormalization}
From (\ref{mainresult8}) we see that for \(N=4\) and \(\nu =\frac{3}{2}\) , \( D= 4-V-E\). This implies that \(\left|\phi\right|^{3}\) has divergences only for \(E=0,1,2,\) and \(3\) and that the Feynman integrals are \emph{superficially convergent} for \(E \geq 4\). From the form of the interaction Lagrangian given by the Hankel contour expansions in (\ref{hankel1}) it is clear that all the odd \(n\)-point Green functions will be \(0\). The \(E=0\) case can be handled via normal ordering. Hence, we only need a mass renormalization for the two-point Green function in the \(\left|\phi\right|^{3}\) theory. So, this theory should be \emph{superrenormalizable}.  For \(\nu=\frac{3}{2}\) in (\ref{normalordergreenfunction2final}) we get 
\begin{equation}
    \left\langle  \phi_{1}\phi_{2}   \right\rangle = \Delta_{12}- \frac{\kappa^{2}}{2} \int d^{4}y \,d^{4}x\, \Delta^{2}_{xy}\,\Delta_{x1}\,\Delta_{y2}+O\left(\kappa^{3}\right)
\end{equation}
The importance of normal ordering can now be seen in a very conspicuous manner. Without normal ordering, we would have gotten ambiguous terms like \(\sqrt{\Delta(0)}\) or \(\left(\Delta(0)\right)^{-\frac{1}{2}}\).   We can now renormalize the \(\left|\phi\right|^{3}\) theory in a straightforward way. The formal (unrenormalized) Lagrangian for this theory is 
\begin{equation}
\mathcal{L}_{U}=\frac{1}{2}\left[(\partial \phi)^{2}-m^{2} \phi^{2}\right]-\frac{\kappa}{3!} \mu^{\epsilon}|\phi|^{3} 
\end{equation}
where the subscript U indicates that it is not renormalized and
\begin{equation}
\epsilon = \frac{4-N}{2}
\end{equation}
Note that $\mu$ is an arbitrary quantity with mass dimensions, which is introduced to keep the coupling constant dimensionless. The 2-point unrenormalized vertex function will be
\begin{equation}
  \Gamma_{U}^{(2)}\left(p^{2}\right)=p^{2}-m^{2}-\frac{\kappa^{2}}{2} \int \frac{d^{N} k}{(2 \pi)^{N}} \frac{1}{\left(k^{2}+m^{2}+i \eta\right)\left[(p-k)^{2}+m^{2}+i \eta\right]}
\end{equation}
Using standard results from dimensional regularization \cite{Sterman1994} we get
\begin{equation}
  \Gamma_{U}^{(2)}\left(p^{2}\right)=p^{2}-m^{2}-\frac{\kappa^{2}}{2} \frac{1}{(4 \pi)^{2}}\left\{\frac{1}{\epsilon}-\gamma+2+\frac{4 \pi \mu^{2}}{m^{2}}-\left(\sqrt{1-\frac{4 m^{2}}{p^{2}}}\right) \ln \left[\frac{\left(\sqrt{1-\frac{4 m^{2}}{p^{2}}}\right)+1}{\left(\sqrt{1-\frac{4 m^{2}}{p^{2}}}\right)-1}\right]\right\}+O(\epsilon)
\end{equation}
where \(\gamma \approx 0.57721 \ldots\) is the Euler-Mascheroni constant. Thus the renormalized Lagrangian will be
\begin{equation}
  \mathcal{L}_{R}=\frac{1}{2}\left[(\partial \phi)^{2}-m_{R}^{2} \phi^{2}\right]-\frac{\kappa}{3!} \mu^{\epsilon}|\phi|^{3}-\frac{1}{2} \delta m^{2} \phi^{2}
\end{equation}
where \(m_{R}\) is the renormalized mass and
\begin{equation}
\begin{aligned}
  m_{R}^{2} & =m^{2}-\delta m^{2} \\
\delta m^{2} & =\frac{\kappa^{2}}{2} \frac{1}{(4 \pi)^{2}}\left(-\frac{1}{\epsilon}+c_{m}\right)
\end{aligned}
\end{equation}
where \(c_{m}\) is a finite constant as \(\epsilon \rightarrow 0\). We will use the \(\overline{MS}\) scheme \cite{Sterman1994}; this will mean that for \(c_{m}\) we have
\begin{equation}
  c_{m} = \gamma-\ln\left(4\pi\right)
\end{equation}
The mass counterterm gives rise to the following perturbation term
\begin{equation}
  -i(2 \pi)^{4} \delta^{4}\left(p-p^{\prime}\right) \delta m^{2}
\end{equation}
The full renormalized 1PI 2-point vertex function \(\Gamma_{R}^{(2)}\) at \(O\left(\kappa^{2}\right)\) is
\begin{equation}
  \Gamma_{R}^{(2)}=p^{2}-m_{R}^{2}-\frac{\kappa^{2}}{(2)(4 \pi)^{2}}\left\{2+\ln \left(\frac{\mu^{2}}{m_{R}^{2}}\right)-\left(\sqrt{1-\frac{4 m_{R}^{2}}{p^{2}}}\right) \ln \left[\frac{\left(\sqrt{1-\frac{4 m_{R}^{2}}{p^{2}}}\right)+1}{\left(\sqrt{1-\frac{4 m_{R}^{2}}{p^{2}}-1}\right)}\right]\right\}+O(\epsilon)
\end{equation}
As previously argued, the only renormalization needed for the \(\left|\phi\right|^{3}\) theory in \(4\) dimensions is a mass renormalization for the two-point Green function. Hence, the formal renormalization of this theory in \(4\) spacetime dimensions is now complete. It should be noted that unlike the regular \(\phi^{3}\) theory which does \emph{not} have a stable ground state \cite{baym1960}, the nonpolynomial theory \(\left|\phi\right|^{3}\) in our formulation does have a proper ground state as can be seen by looking at the form of the interaction Lagrangian. 
\section{Verification in Lower-Dimensions}
We will now verify our results by checking how it appears in spacetime dimensions \(0\) and \(1\). In these lower dimensions, we \emph{do not} have to deal with ultraviolet divergences, hence they are a good playground for testing our results.  Let us state the following result for the \(N\)-dimensional Euclidean scalar Feynman propagator in position space
\begin{equation}
\label{propagator1}
\Delta\left(x_{1}-x_{2}\right) = \frac{1}{\left(2\pi\right)^{\frac{N}{2}}}\, \left|x_{1}-x_{2}\right|^{1-\frac{N}{2}}\,K_{1-\frac{N}{2}}\left(\left|x_{1}-x_{2}\right|\right)
\end{equation}
For \(\Delta(0)\) we have 
\begin{equation}
\label{propagator2}
\Delta(0) = \frac{1}{\left(4\pi\right)^{\frac{N}{2}}} \,\Gamma\left(1-\frac{N}{2}\right)
\end{equation}
We will set the \emph{bare unrenormalized} mass \(m\) equal to \(1\) throughout this section for simplicity. For \(N=0\) we can calculate the first-order shift of the ground state energy using (to first order in \(\kappa\)) 
\begin{align}
\label{calculation1}
 \Delta E &= \frac{\int_{-\infty}^{\infty}dx\, e^{-\frac{x^{2}}{2}}\,\kappa x^{2\nu}}{\int_{-\infty}^{\infty} dx\, e^{-\frac{x^{2}}{2}}}  \nonumber \\
  &= \frac{\kappa}{\sqrt{\pi}}\,  2^{\nu}\, \Gamma\left(\nu+\frac{1}{2}\right)
\end{align}
This is in conformity with our result from \cite{preprint} once we note that \(\Delta(0) =1\) for \(N=0\). For \(N=1\) we can likewise calculate the first-order shift in the ground-state energy through 
\begin{equation}
\Delta E = \frac{\int_{-\infty}^{\infty}dx\, e^{-x^{2}}\,\kappa x^{2\nu}}{\int_{-\infty}^{\infty} dx\, e^{-x^{2}} }  
\end{equation}
The steps are similar to (\ref{calculation1}), except now we will end up with 
\begin{align}
\label{calculation2}
\Delta E &=  \frac{\kappa}{\sqrt{\pi}} \int_{0}^{\infty} du\, u^{-\frac{1}{2}} e^{-u} u^{\nu} \\
&= \frac{\kappa}{\sqrt{\pi}} \, \Gamma\left(\nu+\frac{1}{2}\right)
\end{align}
By using \(\Delta(0)=\frac{1}{2}\) for \(N=1\) (see (\ref{propagator2}) above), we see that this agrees with our result from \cite{preprint} to \(O\left(\kappa\right)\). To verify our result for the \(2\)-point Green function for \(N=0\) we must calculate
\begin{align}
 D_{2} = \frac{\int_{0}^{\infty} dx\, e^{-\frac{x^{2}}{2}}\,x^{2}\left(1-\kappa x^{2\nu}\right)}{\int_{0}^{\infty} dx\, e^{-\frac{x^{2}}{2}}\,\left(1-\kappa x^{2\nu}\right)} &= \frac{\sqrt{\pi}-\kappa 2^{\nu+1} \Gamma\left(\nu+\frac{3}{2}\right)}{\sqrt{\pi}-\kappa 2^{\nu} \Gamma\left(\nu+\frac{1}{2}\right)} \nonumber\\
&= \frac{1}{1+\kappa \frac{\nu}{\sqrt{\pi}} 2^{\nu+1} \,\Gamma\left(\nu+\frac{1}{2}\right)   } 
\end{align}
This is in exact agreement with our result in (\ref{2pointgeneral1}) to \(O\left(\kappa\right)\). For \(N=1\) we must calculate the correction to the first excited state of the harmonic-oscillator Hamiltonian \(H\). The first excited state has the wavefunction (in suitable units with \(m=1\)) \(x e^{-\frac{x^{2}}{2}}\) with eigenvalue \(\frac{3}{2}\). So the time-independent Schrodinger equation 
\begin{equation}
  H\psi = E\psi
\end{equation}
must be solved where
\begin{equation}
\begin{aligned}
\psi(x) &= x e^{-\frac{x^{2}}{2}} +\kappa \psi_{1}(x) +O\left(\kappa^{2}\right) \\
 E &= \frac{3}{2}+\kappa\, E_{1} +O\left(\kappa^{2}\right)
\end{aligned}
\end{equation}
Therefore we have
\begin{equation}
\left(-\frac{1}{2}\frac{d^{2}}{dx^{2}}+\frac{x^{2}}{2}+\kappa x^{2\nu}\right) \left(x e^{-\frac{x^{2}}{2}} +\kappa \psi_{1}(x)\right) = \left(\frac{3}{2}+\kappa\, E_{1}\right)\left(x e^{-\frac{x^{2}}{2}} +\kappa \psi_{1}(x)\right)
\end{equation}
To \(O\left(\kappa\right)\) we get
\begin{equation}
-\frac{1}{2}\frac{d^{2}\psi_{1}(x)}{dx^{2}} +\frac{x^{2}}{2}\psi_{1}(x) - \frac{3}{2}\psi_{1}(x) = \left(E_{1}-x^{2\nu}\right)xe^{-\frac{x^{2}}{2}}
\end{equation}
Since we want to solve for \(E_{1}\) we substitute \(\psi_{1}(x) =xe^{-\frac{x^{2}}{2}}f(x)\)
\begin{equation}
-\frac{x}{2}f''(x)-\frac{1}{2} \left(2-2x^{2}\right)f'(x) +x^{2\nu+1} -\frac{1}{2}\left(x^{3}-3x\right)f(x) +\frac{1}{2} x^{3}f(x) =\frac{3}{2} xf(x) +xE_{1} 
\end{equation}
We now multiply throughout by the integrating factor \(xe^{-x^{2}}\)
\begin{equation}
\begin{aligned}
 & x^{2}e^{-x^{2}} f''(x) +2xe^{-x^{2}}f'(x) -2x^{3}e^{-x^{2}}f'(x)=-2E_{1}x^{2}e^{-x^{2}}+2x^{2\nu+2} e^{-x^{2}} \\
 &\frac{d}{dx}\left(x^{2}e^{-x^{2}}f'(x)\right) = \left(-2E_{1}x^{2}+2x^{2\nu+2}\right)e^{-x^{2}}
\end{aligned}
\end{equation}
We now integrate both sides from \(-\infty\) to \(+\infty\) to get 
\begin{equation}
\label{calculation3}
\begin{aligned}
&\int_{-\infty}^{\infty} dx\, \left(-2E_{1}x^{2}+2x^{2\nu+2}\right) \,e^{-x^{2}} =0\\
&E_{1}= \frac{\int_{-\infty}^{\infty} dx\, x^{2\nu+1}\,e^{-x^{2}}}{\int_{-\infty}^{\infty} dx\, x^{2}\,e^{-x^{2}}   } = \frac{2\,\Gamma\left(\nu+\frac{3}{2}\right)}{\sqrt{\pi}}
\end{aligned}
\end{equation}
The \emph{renormalized mass} \(m_{R}\) is given by 
\begin{equation}
m_{R} = \left(\text{energy of \nth{1} excited state}\right) - \left(\text{energy of ground state}\right)
\end{equation}
Using (\ref{calculation2}) and (\ref{calculation3}) we have 
\begin{equation}
\begin{aligned}
m_{R} &= \left(\frac{3}{2}+\kappa\, \frac{2\Gamma\left(\nu+\frac{3}{2}\right)}{\sqrt{\pi}}\right) - \left(\frac{1}{2}+\kappa\,\frac{\Gamma\left(\nu+\frac{1}{2}\right)}{\sqrt{\pi}}\right)\\
&= 1+\kappa \, \frac{2\nu \,\Gamma\left(\nu+\frac{1}{2}\right)}{\sqrt{\pi}} +O\left(\kappa^{2}\right)
\end{aligned}
\end{equation}
Therefore
\begin{equation}
  m^{2}_{R} = 1+ \kappa\,\frac{4\nu \Gamma\left(\nu+\frac{1}{2}\right)}{\sqrt{\pi}} +O\left(\kappa^{2}\right)
\end{equation}
which agrees perfectly with (\ref{2pointgeneral1}) for \(N=1\) where \(\Delta(0)=\frac{1}{2}\). 
\section{Conclusion}
The results in this paper show that a wide class of hitherto unknown QFTs  might be viable as physical theories (or at least as interesting, new toy models). A complete investigation of the renormalization properties of these nonpolynomial theories is currently underway \cite{CurtrightEuclidean}. It would also be interesting to see how these nonpolynomial theories (especially the inverse-power theories) behave in \(2\) dimensions \cite{Coleman1975} \cite{CurtrightLiouville} \cite{CurtrightThorn1982}. Embedding these theories into fermionic and gauge theories may prove fruitful. 
\appendix\section{Integral Representations}\label{appA}
\begin{figure}[h!]\hspace{-3.5cm}
\hspace*{4cm} 
  \includegraphics[width=0.4\linewidth]{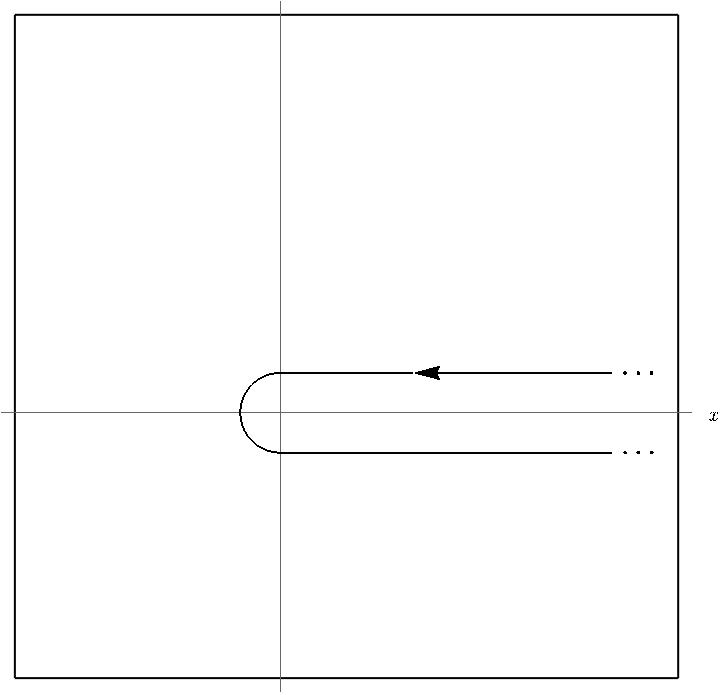}
  \caption{Hankel Contour, with counterclockwise orientation}
  \label{fig:hankel1}
\end{figure}

In \cite{preprint} it was shown how an arbitrary power of the quantum scalar field can be written as 
\begin{equation}
\label{hankel1}
\boxed{\left(\phi^{2}\right)^{\nu} = \frac{i \, \Gamma\left(1+\nu\right)}{2\pi} \oint_{C} ds\, \left(-s\right)^{-1-\nu} \, e^{-s\phi^{2}}}
\end{equation}
where the region of integration \(C\) is the \emph{Hankel contour} whose orientation is as given in FIG. \ref{fig:hankel1}.
Another contour integral expansion that is applicable only for \emph{positive integral} powers is 
\begin{equation}
\label{rep1}
\phi^{n}(x) = \frac{\left(-i\right)^{n} n!}{2\pi i} \oint_{|\lambda|=1} \, \frac{d\lambda}{\lambda^{n+1}} \, e^{i\lambda \phi(x)} \quad \left(n \in \mathbb{Z}_{\geq 0}\right)
\end{equation}
The \emph{Hubbard-Stratonovich Identity} is given by 
\begin{equation}
\label{hubbard1}
e^{-s\phi^{2}} = \int_{-\infty}^{\infty} \frac{d\sigma}{\sqrt{4\pi s}}\, \exp\left(-\frac{\sigma^{2}}{4s}+i\sigma\phi\right) \quad \left(s>0\right)
\end{equation}
An integral identity, sometimes called \emph{Schwinger-parametrization}, that is useful for inverse powers is 
\begin{equation}
\label{rep2}
\frac{1}{(\phi)^{\nu}}  = \frac{1}{\Gamma(\nu)} \int^{\infty}_{0} du \, u^{\nu-1} \, e^{-u\phi}
\end{equation}
In \cite{preprint} it was shown that 
\begin{equation}
\label{preprint1}
  \frac{i\,\Gamma\left(1+\nu\right)}{2\pi} \oint_{C} ds\, \frac{\left(-s\right)^{-1-\nu}}{\sqrt{1+2s\Delta(0)}} = \kappa\,\frac{2^{\nu}\, \Gamma\left(\nu+\frac{1}{2}\right) \,\Delta^{\nu}(0) }{\sqrt{\pi}} 
\end{equation}
By taking appropriate derivatives we can obtain a very useful generalization 
\begin{equation}
\label{appendix2b}
\begin{aligned}
  \frac{i\,\Gamma\left(1+\nu\right)}{2\pi} \oint_{C} ds\, \frac{\left(-s\right)^{-1-\nu} \, s^{m}}{\left(1+2s\Delta(0)\right)^{m+\frac{1}{2}}} = \frac{\left(-1\right)^{m}\,2^{\nu}\,\Gamma\left(\nu+\frac{1}{2}\right)\,\Gamma\left(1+\nu\right) \, \Delta^{\nu-m}(0)   }{\sqrt{\pi}\, \left(2m-1\right)!! \, \Gamma\left(\nu-m+1\right)}
\end{aligned}
\end{equation}
Another representation that we will need is 
\begin{equation}
\label{appendix2}
\begin{aligned}
&\frac{1}{\sqrt{\left(1+2s\Delta(0)\right)\left(1+2t\Delta(0)\right)-4st\Delta^{2}_{xy}   }} \\
&= \frac{1}{\sqrt{\left(1+2s\Delta(0)\right)\left(1+2t\Delta(0)\right)}} \, \frac{1}{\sqrt{1-\frac{4st\Delta^{2}_{xy}}{\left(1+2s\Delta(0)\right)\left(1+2t\Delta(0)\right)}}} \\
&= \sum_{m=0}^{\infty}\,\frac{2^{m} \Delta^{2m}_{xy}\left(2m-1\right)!!}{m!}\left(\frac{s^{m}}{\left(1+2s\Delta(0)\right)^{m+\frac{1}{2}}}\right) \left(\frac{t^{m}}{\left(1+2t\Delta(0)\right)^{m+\frac{1}{2}}}\right) .
\end{aligned}
\end{equation}
Similarly it can be shown 
\begin{equation}
\label{appendix2a}
\begin{aligned}
&\frac{1}{\left(\left(1+2s\Delta(0)\right)\left(1+2t\Delta(0)\right)-4st\Delta^{2}_{xy}\right)^{\frac{3}{2}}} \\
&= \sum_{m=0}^{\infty} \frac{2^{m}\,\Delta^{2m}_{xy} \, \left(2m+1\right)!!}{m!} \, \left(\frac{s^{m}}{\left(1+2s\Delta(0)\right)^{m+\frac{3}{2}}}\right)\, \left(\frac{t^{m}}{\left(1+2t\Delta(0)\right)^{m+\frac{3}{2}}}\right) .
\end{aligned}
\end{equation}

\subsection{VEV of products of exponentials}
\begin{equation}
\label{vevgaussians2}
\left\langle e^{i\sigma \phi_{x}} e^{i\tau \phi_{y}}    \right\rangle_{0} =  \exp\left(-\frac{1}{2}\left(\tau^{2}+\sigma^{2}\right)\Delta(0)- \sigma\tau\Delta_{xy} \right)
\end{equation}
\begin{proof}
We can rewrite the VEV using a source term 
\begin{equation}
\begin{aligned}
      \left\langle e^{i\sigma\phi_{x} + i\tau\phi_{y}}  \right\rangle_{0} =  \left\langle e^{i\int d^{N}z \, J_{z} \,\phi_{z}  }  \right\rangle_{0}
\end{aligned}
\end{equation}
where \(J(z) = \sigma \delta^{(N)}_{xz}+ \tau \delta^{(N)}_{yz}\). For free fields with a source term \(J\) we have
\begin{equation}
    \left\langle e^{i\int d^{N}z \, J(z)\phi(z)  }  \right\rangle = e^{-\frac{1}{2}\int dz\,dz'\, J(z)\Delta_{zz'}}
\end{equation}
For our  form of \(J\) we have
\begin{equation*}
    -\frac{1}{2}\int d^{N}z\,d^{N}z'\, J_{z} \,\Delta_{zz'}\,J_{z'} = -\frac{1}{2}\left\{\sigma^{2}\Delta\left(0\right)+\tau^{2}\Delta\left(0\right) + 2\sigma\tau\Delta_{xy} \right\}
\end{equation*}
This means that 
\begin{equation}
  \left\langle e^{i\sigma \phi_{x}} e^{i\tau \phi_{y}}    \right\rangle_{0} = \exp\left(-\frac{1}{2}\left(\tau^{2}+\sigma^{2}\right)\Delta(0)- \sigma\tau\Delta_{xy} \right)
\end{equation}
\end{proof} 
We can now see that for a product of \(n\) exponentials we would get 
\begin{equation}
\label{vevexpn}
 \left\langle \prod_{j=1}^{n} e^{i\sigma_{j}\phi\left(y_{j}\right)}\right\rangle_{0} = \exp\left( -\Delta(0)\sum_{j=1}^{n} \sigma^{2}_{j} -\sum_{j <k} \sigma_{j}\sigma_{k} \Delta\left(y_{j}-y_{k}\right) \right).
\end{equation}
\section{Vacuum Expectation Value(VEV) of a variety of field functions}\label{appB}
\begin{equation}
\label{appendix1}
\begin{aligned}
\left\langle \phi_{1}\phi_{2}e^{-s\phi^{2}_{y}} \right\rangle_{0} &= \frac{\left(-i\right)^{2}}{\left(2\pi i\right)^{2}} \oint_{\left|\lambda\right|=\left\|\tau\right\|=1} \frac{d\lambda\,d\tau}{\lambda^{2}\,\tau^{2}}\, \int_{-\infty}^{\infty} \frac{d\sigma}{\sqrt{4\pi s}} \, e^{-\frac{\sigma^{2}}{4s}} \, \left\langle  e^{i\lambda \phi_{2}}\,e^{i\tau \phi_{y}}\,e^{i\sigma\phi_{y}}  \right\rangle_{0} \\
&= \left.\frac{\left(-i\right)^{2}}{\left(2\pi i\right)^{2}} \left(2\pi i\right)^{2} \, \frac{d}{d\lambda}\frac{d}{d\tau} \int_{-\infty}^{\infty} \frac{d\sigma}{\sqrt{4\pi s}} \, e^{-\frac{\Delta(0)}{2} \left(\tau^{2}+\sigma^{2}+\lambda^{2}\right)} \,   e^{  -\sigma\tau\Delta(0) -\lambda\tau\Delta_{y2}-\lambda\sigma\Delta_{y2}     }     \right|_{\lambda=\tau=0} \\
&= \frac{\Delta_{y2}}{\left(1+2s\Delta(0)\right)^{\frac{3}{2}}}
\end{aligned}
\end{equation}
where we used (\ref{vevexpn}). Using (\ref{hubbard1}) and (\ref{rep1}) we also get

\begin{equation}
\label{appendix3}
\begin{aligned}
\left\langle  \phi_{y}\phi_{2}\phi_{3}\phi_{4}\,e^{-s\phi^{2}_{y}}    \right\rangle_{0} = -\frac{6s \, \Delta_{y2}\,\Delta_{y3}\,\Delta_{y4}}{\left(1+2s\Delta(0)\right)^{\frac{5}{2}}} + \frac{\Delta_{y4}\,\Delta_{23}}{\left(1+2s\Delta(0)\right)^{\frac{3}{2}}} +  \frac{\Delta_{y3}\,\Delta_{24}}{\left(1+2s\Delta(0)\right)^{\frac{3}{2}}} + \frac{\Delta_{y2}\,\Delta_{34}}{\left(1+2s\Delta(0)\right)^{\frac{3}{2}}}  
\end{aligned}
\end{equation}
Similarly 
\begin{equation}
\label{appendix4}
  \left\langle   e^{-t\phi^{2}_{x}} \,\phi_{x} \,     \phi^{2n-2m-1}_{y} \right\rangle_{0} = \frac{  \Gamma\left(n-m+\frac{1}{2}\right)\,2^{2n-2m-1} }{\sqrt{\pi}\, \left(1+2t\Delta(0)\right)^{\frac{3}{2}}} \,\left(\frac{2t\Delta^{2}(0)-2t\Delta^{2}_{xy}+\Delta(0)}{2+4t\Delta(0)}\right)^{n-m-1}\, \Delta_{xy}
\end{equation}
\begin{equation}
\label{appendix5}
\left\langle  e^{-s\phi^{2}_{x}}e^{-t\phi^{2}_{y}} \phi_{x}\phi_{y}   \right\rangle_{0} = \frac{\Delta_{xy}}{\left(\left(1+2s\Delta\right)\left(1+2t\Delta\right)-4st\Delta^{2}_{xy}\right)^{\frac{3}{2}}}
\end{equation}

\section{Veltman's Rule}\label{appC}
In dimensional regularization, \emph{Veltman's rule} is  
\begin{equation}
\label{veltman}
  \int \frac{d^{N}p}{\left(2\pi\right)^{N}} \, \left(p^{2}\right)^{\lambda} = 0  
\end{equation}
where \(\lambda, N \in \mathbb{C}\). For more details see chapter \(8\) in \cite{Kleinert2001}.


\bibliography{references}

\end{document}